\documentclass{article}
\usepackage{spconf,amsmath,graphicx}
\usepackage{multirow}
\usepackage{booktabs}
\usepackage{hyperref}

\title{HyperGANStrument: Instrument Sound Synthesis and Editing with Pitch-Invariant Hypernetworks}
\name{Zhe Zhang and Taketo Akama}
\address{Sony Computer Science Laboratories, Tokyo, Japan}
\begin{document}
%
\maketitle
\begin{abstract}

GANStrument, exploiting GANs with a pitch-invariant feature extractor and instance conditioning technique, has shown remarkable capabilities in synthesizing realistic instrument sounds. To further improve the reconstruction ability and pitch accuracy to enhance the editability of user-provided sound, we propose HyperGANStrument, which introduces a pitch-invariant hypernetwork to modulate the weights of a pre-trained GANStrument generator, given a one-shot sound as input. The hypernetwork modulation provides feedback for the generator in the reconstruction of the input sound. In addition, we take advantage of an adversarial fine-tuning scheme for the hypernetwork to improve the reconstruction fidelity and generation diversity of the generator. Experimental results show that the proposed model not only enhances the generation capability of GANStrument but also significantly improves the editability of synthesized sounds. Audio examples are available at the online demo page\footnote{\label{demo}\url{https://noto.li/MLIuBC}}.

\end{abstract}
\section{Introduction}
\label{sec:intro}

Instrument sound synthesis is an important and interesting topic in both music technique research and industry. Traditional methods, such as additive, subtractive, and physical modeling synthesis, have provided the foundation for creating a wide variety of sounds. However, these methods often have limitations on wide-range generation fidelity and timbre editing flexibility. With the advent of deep learning and its success in generative modeling, there has been a growing interest in leveraging these models for instrument sound synthesis. In this paper, we tackled instrumental sound synthesis and editing given a one-shot sound input, realizing a deep neural sampler with high-fidelity and diverse generation ability.

Traditional samplers often record and playback with audio effects. However, it is difficult to create new timbres or mix multiple timbres intelligently. Through the use of deep generative models and latent space exploration, recent audio synthesis models are able to generate and mix diverse timbres flexibly \cite{narita_ganstrument_2023,shan_differentiable_2022,nistal_drumgan_2022,engel_ddsp_2020,luo_learning_2019,engel_gansynth_2018,engel_neural_2017}. Especially, GANStrument \cite{narita_ganstrument_2023}, taking advantages of StyleGAN2 \cite{karras_analyzing_2020} with a pitch-invariant feature extractor, has demonstrated its capability to produce a wide range of realistic and novel instrument timbres. However, the generation quality can be sometimes degraded given the input of complex timbre or non-instrumental sounds.
In real-world scenarios, it is important for neural synthesizers to generate diverse high-quality sounds with accurate pitch.

Inspired by HyperStyle \cite{alaluf_hyperstyle_2022}, addressing the challenge of inversion of real images into pre-trained generator's latent space by introducing hypernetworks, we propose HyperGANStrument, a novel neural synthesizer that integrates the principles of GANStrument with the hypernetwork-based inversion techniques \cite{ha_hypernetworks_2016}. Aiming at enhancing pitch accuracy and sound fidelity of deep samplers, we present a pitch-invariant hypernetwork to modulate the weights of the pre-trained generator. In addition, we leverage a conditional adversarial fine-tuning scheme to train the hypernetwork. We demonstrate that our model is lightweight and efficient, which is crucial for real-world applications as musical instruments.

\begin{figure*}[htb]
\centering
\includegraphics[width=1.8\columnwidth]{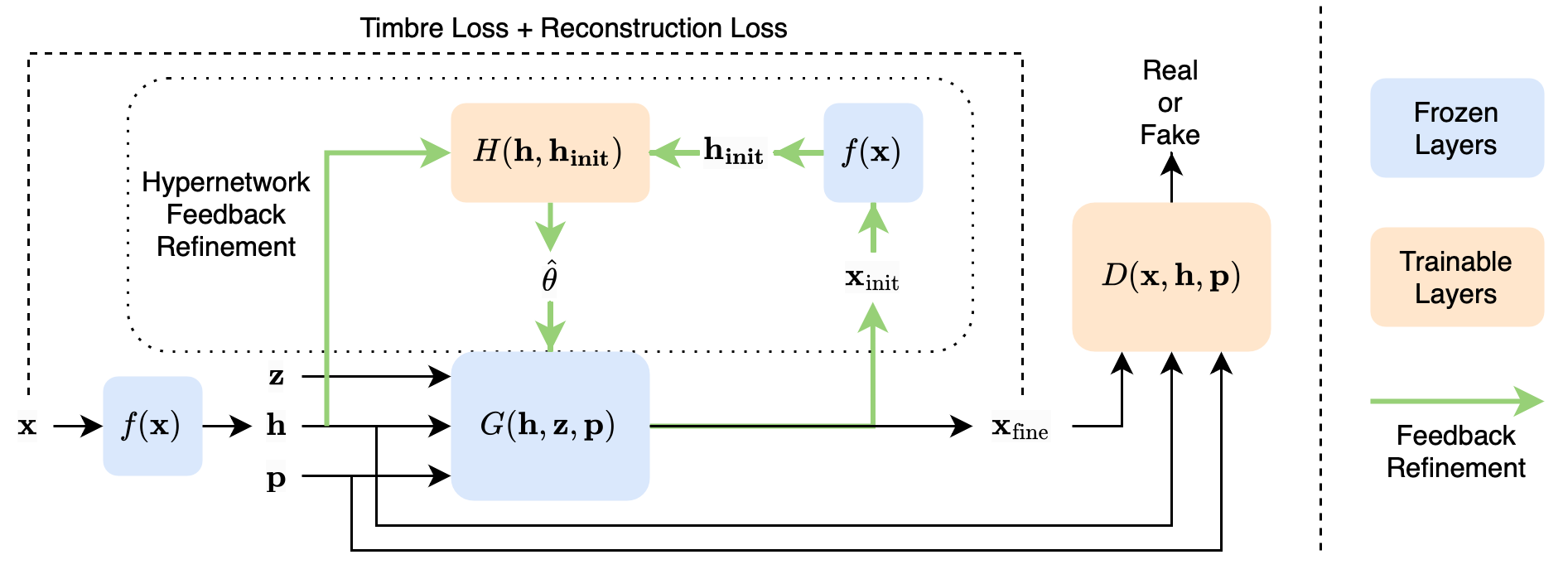}
\caption{An overview of HyperGANStrument.}
\label{fig:overview}
\end{figure*}

\section{Related Work}

GANStrument \cite{narita_ganstrument_2023} leverages the architecture of StyleGAN2 \cite{karras_analyzing_2020} and instance conditioning to generate instrument sounds, which overcomes some limitations of previous neural synthesizers like WaveNet \cite{oord_wavenet_2016}, GANSynth \cite{engel_gansynth_2018}, DDSP \cite{engel_ddsp_2020,shan_differentiable_2022} on generation quality and input editability. RAVE \cite{caillon_rave_2021} exploits VAE and GANs towards real-time and high-quality timbre transfer. Style transfer is also studied for timbre control in \cite{wu_transplayer_2023}. Nevertheless, there is still room for enhancement for reflecting fine-grained sound texture of input and robust pitch control.

GAN inversion \cite{zhu_generative_2016} aims to obtain a latent code of input for reconstruction. An encoder can be trained to learn a mapping from an image to its latent representation \cite{alaluf_restyle_2021,tov_designing_2021,wang_high-fidelity_2022}. Hypernetwork \cite{alaluf_hyperstyle_2022,dinh_hyperinverter_2022} is another type of approach addressing the challenge of mitigating the trade-off between reconstruction and editability

\section{Methods}

In this section, we introduce the details of HyperGANStrument and its training pipeline. As shown in Fig. \ref{fig:overview}, the input waveform is first transformed into its mel-spectrogram $\mathbf{x}$, and its feature $\mathbf{h}$ is extracted with the pre-trained feature extractor $f(\mathbf{x})$ from GANStrument. Then, it is fed into the pre-trained GANStrument generator $G$ together with pitch $\mathbf{p}$ and noise $\mathbf{z}$ to reconstruct a mel-spectrogram $\mathbf{x}_{\rm{init}}$. The hypernetwork is defined to predict the offsets of the weights for the generator. With the feedback from the hypernetwork, the generator synthesizes refined mel-spectrogram $\mathbf{x}_{\rm{fine}}$.

\subsection{Hypernetwork}

Inspired by HyperStyle, which aims to learn to efficiently optimize the generator for a given target image, we propose the hypernetwork $H$ to learn to improve the pre-trained generator $G(\mathbf{h}, \mathbf{z}, \mathbf{p};\theta)$ in both generation quality and pitch accuracy. HyperStyle takes original and reconstruction images as the input of the hypernetwork, which we find is harmful to the pitch-timbre disentanglement in our task. Instead, our hypernetwork takes pitch-invariant features from the GANStrument feature extractor $\mathbf{h} = f(\mathbf{x})$, given by
\begin{equation}
\hat{\theta}=H\left(\mathbf{h}, \mathbf{h}_{\rm{init}}\right),
\end{equation}
where $\mathbf{h}_{\rm{init}}$ is the extracted feature of the initial reconstruction $\mathbf{x}_{\rm{init}}$ by the generator, as illustrated in Fig. \ref{fig:overview}.

After careful design and experiments of the architecture of $H$, we firstly use a shallow MLP network consisting of two linear layers to map our input features into dimensions of $1\times1\times512$. Following the MLP layer, we then adopt the Refinement Blocks and the Shared Refinement Blocks from \cite{alaluf_hyperstyle_2022} to predict per-layer offsets for the parameters of the generator.

\subsection{Feedback Refinement}

After predicting the offsets from the initial reconstruction, the hypernetwork gives feedback to the generator to update its parameters. Then, the generator is able to generate refined mel-spectrograms with the updated parameters, given by
\begin{equation}
    {\mathbf{x}_\mathrm{fine}} = G\left(\mathbf{h}, \mathbf{z}, \mathbf{p} ; \hat{\theta}\right).
\end{equation}

Then, the hypernetwork is trained to minimize the reconstruction error between the real sound $\mathbf{x}$ and the refined sound $\hat{\mathbf{x}}$, ensuring accurate and efficient inversion, given by
\begin{equation}
    \mathcal{L_{\text{pre}}}\left(\mathbf{x}, G\left(\mathbf{h}, \mathbf{z}, \mathbf{p} ; \hat{\theta}\right)\right),
\end{equation}
where $\mathcal{L_{\text{pre}}}$ is the learning objective for the pre-training stage.

For training the hypernetwork, we adopt the instance conditioning technique \cite{casanova_instance-conditioned_2021} from GANStrument. Additionally, we set a hyperparameter $p_{\mathrm{recon}}$ as the probability of using the pitch of the input sound instead of the sampled k-nearest neighbor (KNN) pitch by instance conditioning. Thus, the hypernetwork is trained by timbre loss and reconstruction loss to make $\mathbf{x}_{\rm{fine}}$ closely matches $\mathbf{x}$ in a pitch-invariant way, where timbre loss is the $L2$ distance between the features extracted by the pitch-invariant feature extractor $f(\mathbf{x})$ and the reconstruction loss is the $L2$ distance between the mel-spectrograms. Formally, the loss objective is given by
\begin{equation}
    \mathcal{L}_{\text{pre}} = \lambda_{\mathrm{timbre}}\mathcal{L}_{2}\left(\mathbf{h}, \mathbf{h}_{\rm{fine}}\right) + \lambda_{\mathrm{recon}}\mathcal{L}_{2}\left(\mathbf{x}, \mathbf{x}_{\rm{fine}}\right),
\end{equation}
where $\lambda_{\mathrm{recon}} = 0$ when not using label pitch. Otherwise, $\lambda_{\mathrm{recon}}=100$ and $\lambda_{\mathrm{timbre}} = 1$ during pre-training.

\subsection{Conditional Adversarial Fine-tuning}

After the above training of hypernetwork converges, to further enhance the quality and editability of the synthesized sounds, we introduce a conditional adversarial fine-tuning process. Specifically, we introduce the projection discriminator \cite{miyato_cgans_2018} $D(\mathbf{x},\mathbf{p},\mathbf{h})$ from GANStrument to distinguish between real and synthesized sounds while being aware of the pitch-timbre disentanglement. In the adversarial fine-tuning stage, the discriminator and the hypernetwork are jointly trained by
\begin{equation}
\mathcal{L}\left(G\right) = -\log D\left(\mathbf{x}'_{\rm{fine}},\mathbf{p}',\mathbf{h}\right) +  \mathcal{L}_{\text{pre}},
\end{equation}
\begin{equation}
\mathcal{L}\left(D\right) = -\log D\left(\mathbf{x}',\mathbf{p}',\mathbf{h}\right) + \log D\left(\mathbf{x}'_{\rm{fine}},\mathbf{p}',\mathbf{h}\right),
\end{equation}
where $\mathbf{x}'$ and $\mathbf{p}'$ are the sampled mel-spectrogram by KNN sampling of instance conditioning and its pitch label, respectively. $\mathbf{x}'_\mathrm{fine}$ is the refined generator prediction conditioned by $\mathbf{p}'$. As mentioned above, there is a probability of $p_{\mathrm{recon}}$ that the pitch of input sound instead of the sampled pitch is used. In such cases, we have $\mathbf{x}' = \mathbf{x}$ and $\mathbf{p}' = \mathbf{p}$. Moreover, to stabilize the training process, we use a fixed $\mathbf{z} = [0, 0, ..., 0]$ in training and inference stage. The pre-trained generator is frozen in training, thus only the parameters of the hypernetwork and the discriminator are updated. In the adversarial fine-tuning stage, $\lambda_{\mathrm{timbre}}$ is set to 20 and $\lambda_{\mathrm{recon}}$ is set to 200.

By optimizing the hypernetwork with the conditional discriminator, the model is trained to improve the realism of the sound and the effectiveness of the pitch conditioning given timbre feature. This is important for retaining the ability to generate diverse sounds while allowing for accurate control over the pitch. Verified by experiments, this conditional adversarial fine-tuning process further improves the hypernetwork to make the generated sounds both realistic and editable.

\begin{table*}[htb]
\centering
\caption{Evaluation of HyperGANStrument}
\begin{tabular}{c|cc|cc|cc}
\toprule
\multirow{2}{*}{Models} & \multicolumn{2}{c|}{Reconstruction} & \multicolumn{2}{c|}{Synthesis} & \multicolumn{2}{c}{Interpolation} \\
\cmidrule{2-7}
& MSE$\downarrow$ & Pitch$\uparrow$ & FID$\downarrow$ & Pitch$\uparrow$ & FID$\downarrow$ & Pitch$\uparrow$\\
\midrule
GANStrument & 1.79 & 0.90 & 212.3 & 0.87 & 252.2 & 0.88\\
GANStrument Enc. & 1.72 & 0.91 & 253.7 & 0.87  & 260.3 & 0.87\\
HyperGANStrument-Pre & 1.49 & \textbf{0.93} & 242.2 & \textbf{0.91} & 248.3 & \textbf{0.91}\\
\midrule
\textbf{HyperGANStrument (ours)} & \textbf{1.30} & \textbf{0.93} & \textbf{153.1} & 0.90 & \textbf{156.4} & \textbf{0.91}\\
\bottomrule
\end{tabular}
\label{tab:evaluation}
\end{table*}

\section{Evaluation}

\subsection{Experiment Setup}

We trained HyperGANStrument on NSynth dataset \cite{engel_neural_2017}, a comprehensive collection of instrument sounds. The dataset was pre-processed to extract 88 MIDI notes (21-108) and apply amplitude normalization, following the settings of GANStrument and to cover the majority of instrument pitch range. In the evaluation, following GANStrument, we used the NSynth validation dataset. The dimension of mel-spectrogram is $512 \times 512$, derived by an STFT with a Hann window, a 1024 window size, a 64 hop size, a 2048 FFT size, and followed by mel-scale conversion with 512 filter banks. 

We utilized a pre-trained GANStrument model along with its feature extractor as the basis of our HyperGANStrument model. The ADAM optimizer \cite{kingma_adam_2017} with a learning rate of $2.5 \times 10^{-3}$ was used both in pre-training and fine-tuning. The hypernetwork is trained with $\mathcal{L}_{\text{pre}}$ for 200k iterations first, then jointly trained the hypernetwork and the discriminator for 200k iterations. In the adversarial training stage, $R_1$ regularization \cite{karras_analyzing_2020} technique is exploited for the stability of the training process. We keep other settings same with GANStrument for fair comparison. Besides, the probability of using label pitch for training the generator is chosen as $p_{\text{recon}} = 0.2$.

\subsection{Generation Quality}

\begin{figure}[htb]
\centering
\includegraphics[width=0.9\columnwidth]{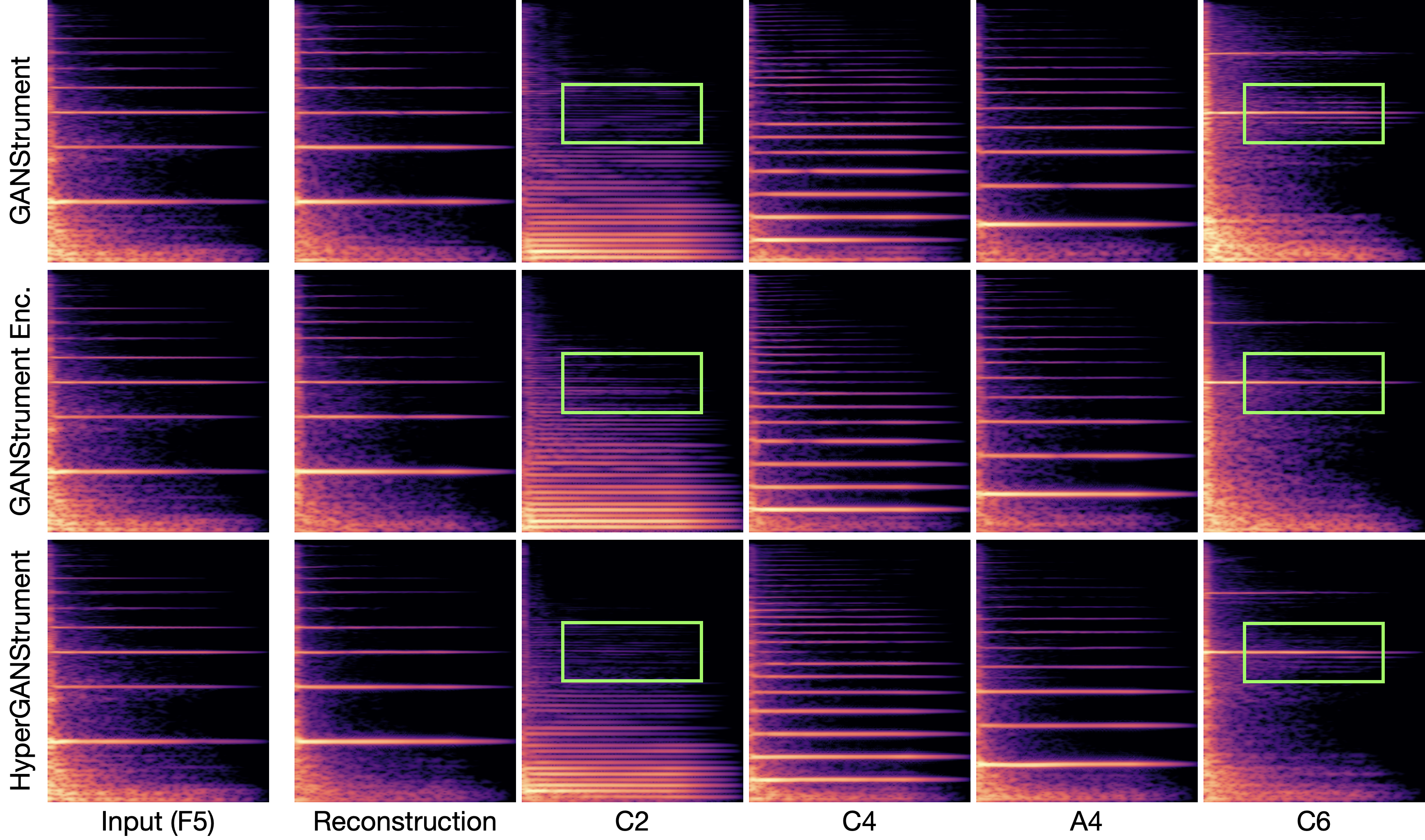}
\caption{Examples of reconstruction and generation.}
\label{fig:generation}
\end{figure}

To assess generation quality and pitch accuracy of HyperGANStrument, we conducted a series of qualitative and quantitative evaluations. In addition to GANStrument, which already outperformed conditional GAN models and their encoder-based inversion models in generation task and sound reconstruction task \cite{narita_ganstrument_2023}, we trained an encoder-based GAN inversion model \cite{alaluf_restyle_2021} $(\mathbf{h}, \mathbf{z}) = E(\mathbf{x})$ for the pre-trained GANStrument model $G$ as another strong baseline, denoted as GANStrument Enc. in Table \ref{tab:evaluation} and Fig. \ref{fig:generation}. The training objective for the encoder is $\min_E\|\mathbf{x} - G(E(\mathbf{x}), \mathbf{p})\|^2 + \lambda\|\mathbf{z}\|^2$, where the second term is a regularization for $\mathbf{z}$ to follow a standard normal distribution. The model is trained for 1,200k iterations to achieve convergence and its best performance. Moreover, we trained a HyperGANStrument-Pre model only with $\mathcal{L}_{\text{pre}}$ for 400k iterations as another ablation model without adversarial fine-tuning. 

We evaluate the models with two kinds of generation tasks, i.e., reconstruction and synthesis. In the reconstruction experiment, we evaluate the faithfulness of the reconstructed sounds by computing the mean square error (MSE) between the features extracted by a pre-trained instrument category classifier adopted from \cite{narita_ganstrument_2023}. In the synthesis experiment, we evaluate the generation ability of the models by conditioning the input timbre with arbitrary MIDI pitch. Fr\'{e}chet inception distance (FID) \cite{heusel_gans_2018} from the instrument category classifier is used to measure the distances between the generated sounds and ground-truth sounds in the timbre feature space. In both experiments, pitch accuracy is also evaluated by a pitch classifier trained on the NSynth dataset.

The left and middle parts of Table \ref{tab:evaluation} show the evaluation results of reconstruction and synthesis. The metrics demonstrate that HyperGANStrument can outperform the baseline models on both sound fidelity and pitch accuracy, confirming the effectiveness of the proposed hypernetwork feedback refinement technique and the adversarial fine-tuning scheme. The notable improvement of pitch accuracy is also crucial for HyperGANStrument to act as a deep neural sampler in real-world music applications.

Fig. \ref{fig:generation} shows the reconstruction and generation results conditioned by different MIDI pitch values from GANStrument, GANStrument Encoder, and HyperGANStrument, given a piano sound as input. We can observe that although GANStrument and its encoder inversion already predicts high-quality mel-spectrograms, sometimes the synthesized sounds tend to be over-harmonized, leading to an unnatural timbre. Instead, HyperGANStrument can generate a more fine-grained sound texture, as shown in the green rectangles. We refer readers to our demo page for audible examples\textsuperscript{\ref{demo}}.

\subsection{Interpolation}

Editability is also an important aspect of HyperGANStrument. We proposed a pipeline of latent space interpolation to mix multiple timbres. Specifically, suppose that we want to mix two timbres from sounds $\mathbf{x}_1$ and $\mathbf{x}_2$, firstly we extract the timbre features by $f(\mathbf{x})$ and interpolate them in timbre space by $\mathbf{h}_{\text{interp}} = \alpha_1\mathbf{h}_{1} + \alpha_2\mathbf{h}_{2}$. Then, we generate the initial reconstruction of the mixed sound by $\mathbf{x}_{\text{init}} = G\left(\mathbf{h}_{\text{interp}}, \mathbf{z}, \mathbf{p}\right)$. Afterwards, this interpolated initial reconstruction was fed into the proposed feedback refinement process to generate the refined sounds by the hypernetwork.

In the quantitative experiments, we randomly interpolate the timbres in each batch and generate the mixed sounds conditioned on arbitrary pitch. Then we assess the FID and the pitch accuracy of the mixed sounds. As shown in the right part of Table \ref{tab:evaluation}, our HyperGANStrument achieves both better FID and pitch accuracy, demonstrating the high-quality generated sounds with accurate pitch from latent space exploration.

\begin{figure}[htb]
\centering
\includegraphics[width=0.9\columnwidth]{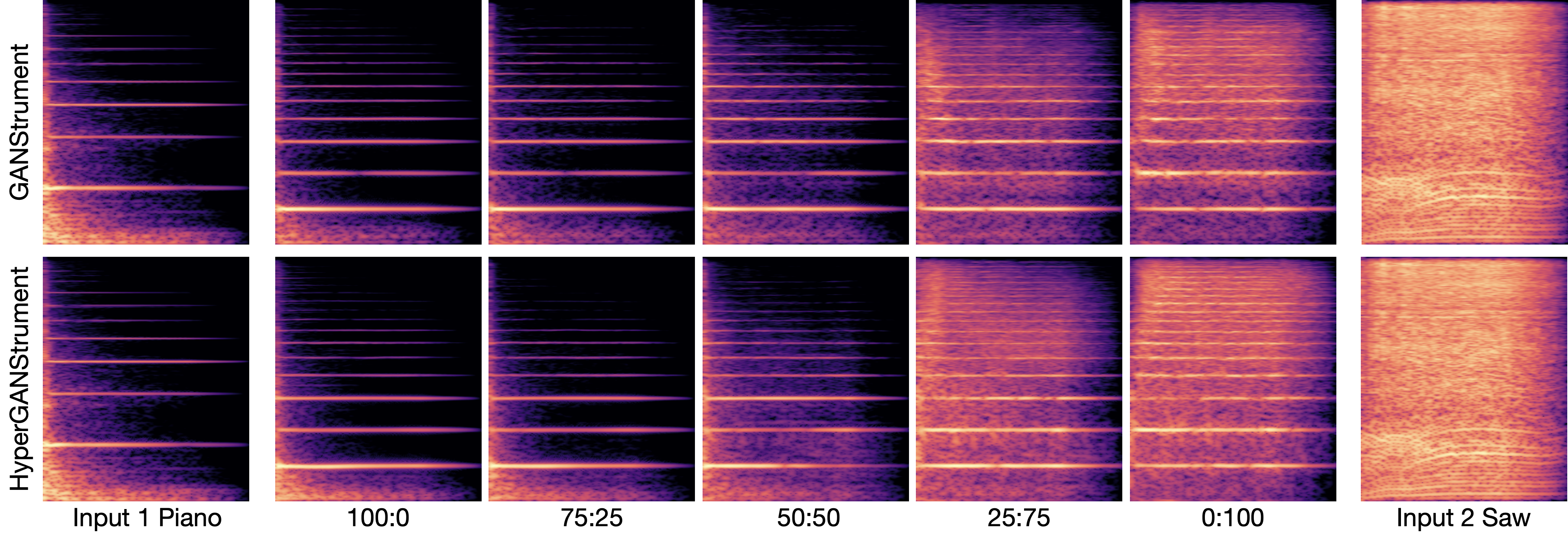}
\caption{Examples of interpolation.}
\label{fig:interpolation}
\end{figure}
Fig. \ref{fig:interpolation} shows an example of latent interpolation. We chose two extremely heterogeneous timbres, piano (F5) and saw sound, as input and interpolated them in the timbre space with various ratios, conditioned on pitch A4. It is worth mentioning that the saw sound is very noisy and totally out of the scope of the musical training data. By watching the mel-spectrogram and listening to the samples, HyperGANStrument is more successful in capturing the timbre feature of the saw sound and mixing it with the harmonious piano sound. This not only verified the strong editability of HyperGANStrument with disentanglement between pitch and timbre but also suggested a better generalization ability to unseen non-instrument sounds. More audio smaples are available at the demo page\textsuperscript{\ref{demo}}.

\subsection{Ablation Study}

We conducted ablation study to confirm the significance of pitch-invariant hypernetwork. A ResNet-based \cite{he_deep_2015} hypernetwork proposed in \cite{alaluf_hyperstyle_2022}, taking mel-spectrograms as input to predict the parameter offsets for the generator, was trained with the same settings. Consequently, the model showed degraded performance in reconstruction, where MSE and pitch accuracy were 3.43 and 0.58, respectively. Moreover, the training process became unstable. We argue that the pitch information in mel-spectrogram corrupted GANStrument's pitch-disentangled timbre space , thus leading to worse results. This demonstrates the necessity of the proposed hypernetwork with pitch-invariant input and training objectives.

\subsection{Efficiency}

Efficiency is crucial for real-time musical instrument synthesis. The highly-pruned hypernetwork in \cite{alaluf_hyperstyle_2022} has $332M$ parameters. Thanks to the feature extractor of GANStrument, we further reduced this number to $273M$ by our pitch-invariant hypernetwork. Moreover, we measured the time to generate a sound sample on an Intel 3.0GHz CPU. On average, the generation time of HyperGANStrument only increased $0.439s$ compared with GANStrument, which preserves the interactive generation time while increasing the sound quality.




\section{Conclusion}

In this paper, we proposed our hypernetwork-based neural synthesizer, HyperGANStrument, to enhance the generation ability and editability of the pre-trained GANStrument model. By training the pitch-invariant hypernetwork with the conditional adversarial fine-tuning pipeline, the generator is able to achieve better reconstruction fidelity, pitch accuracy, and generalization ability with the feedback refinement technique. Experimental results verified the superiority of HyperGANStrument, which will enable musicians to freely explore novel, diverse, and high-quality sound timbres.

\vfill\pagebreak
\bibliographystyle{IEEEbib}
\bibliography{strings,refs}

\end{document}